\documentclass[12pt]{amsart}
\usepackage{amsmath}
\usepackage{amsxtra}
\usepackage{amscd}
\usepackage{amsthm}
\usepackage{amsfonts}
\usepackage{amssymb}
\usepackage{eucal}
\def\({\left(}
\def\){\right)}
\def\[{\left[}
\def\]{\right]}

\newcommand{\nn}{\nonumber}
\newcommand{\bea}{\begin{eqnarray}}
\newcommand{\ena}{\end{eqnarray}}
\def\bel{\begin{eqnarray}}
\def\enl{\end{eqnarray}}
\newcommand{\be}{\begin{eqnarray*}}
\newcommand{\en}{\end{eqnarray*}}
\newcommand{\ba}{\begin{array}}
\newcommand{\ea}{\end{array}}

\newcommand{\bb}{\mathbf{b}}
\newcommand{\bc}{\mathbf{c}}
\newcommand{\C}{{\mathbb C}}

\newcommand{\bbA}{\mathbb{A}}
\newcommand{\bbB}{\mathbb{B}}
\newcommand{\bbC}{\mathbb{C}}
\newcommand{\bbD}{\mathbb{D}}
\newcommand{\bQ}{\mathbf{Q}}

\newcommand{\cP}{\mathcal{P}}

\newcommand{\cR}{\mathcal{R}}
\newcommand{\bX}{\mathbf{X}}

\newcommand{\bT}{\mathbb{T}}

\newcommand{\Ob}{\mbox{\boldmath$\Omega $}}
\newcommand{\ob}{\mbox{\boldmath$\omega $}}
\newcommand{\Obm}{\mbox{\scriptsize\boldmath{$\Omega$}}}
\newcommand{\slt}{\mathfrak{sl}_2}
\newcommand{\slth}{\widehat{\mathfrak{sl}}_2}
\newcommand{\res}{{\rm res}}

\newcommand{\tr}{{\rm tr}}

\newcommand{\Tr}{{\rm Tr}}

\def\bS{\mathbf{S}}

\newenvironment{tenumerate}{
  \begin{enumerate}
  
  }{\end{enumerate}}
\newcommand{\bi}{\begin{tenumerate}}
\newcommand{\ei}{\end{tenumerate}}
\newcommand{\isoto}[1][]%
{{\mathop{\buildrel{\sim}\over\longrightarrow}\limits_{#1}}}

\newcommand{\al}{\alpha}

\newcommand{\z}{\zeta}

\numberwithin{equation}{section}
\newtheorem{thm}{Theorem}[section]

\newtheorem{lem}[thm]{Lemma}

\newtheorem{definition}[thm]{Definition}

\begin{document}

\begin{title}[Grassmann structure in the XXZ model]
{Hidden Grassmann structure in the XXZ model}
\end{title}
\date{\today}
\author{H.~Boos, M.~Jimbo, T.~Miwa, F.~Smirnov and Y.~Takeyama}
\address{HB: Physics Department, University of Wuppertal, D-42097,
Wuppertal, Germany\footnote{
on leave of absence from 
Skobeltsyn Institute of Nuclear Physics, 
MSU, 119992, Moscow, Russia
}}\email{boos@physik.uni-wuppertal.de}
\address{MJ: Graduate School of Mathematical Sciences, The
University of Tokyo, Tokyo 153-8914, Japan}\email{jimbomic@ms.u-tokyo.ac.jp}
\address{TM: Department of Mathematics, Graduate School of Science,
Kyoto University, Kyoto 606-8502, 
Japan}\email{tetsuji@math.kyoto-u.ac.jp}
\address{FS\footnote{Membre du CNRS}: Laboratoire de Physique Th{\'e}orique et
Hautes Energies, Universit{\'e} Pierre et Marie Curie,
Tour 16 1$^{\rm er}$ {\'e}tage, 4 Place Jussieu
75252 Paris Cedex 05, France}\email{smirnov@lpthe.jussieu.fr}
\address{YT:  Institute of Mathematics, 
Graduate School of Pure and Applied Sciences, Tsukuba University, 
Tsukuba, Ibaraki 305-8571, Japan}
\email{takeyama@math.tsukuba.ac.jp}

\begin{abstract}
{}For the critical XXZ model, 
we consider the space $\mathcal{W}_{[\al]}$ of operators which are
products of local operators with a disorder operator.
We introduce two anti-commutative family of operators 
$\bb(\z),\bc(\z)$ which act on $\mathcal{W}_{[\al]}$.   
These operators are constructed as traces over 
representations of the $q$-oscillator algebra,    
in close analogy with Baxter's $Q$-operators.  
We show that the vacuum expectation values 
of operators in $\mathcal{W}_{[\al]}$ can be expressed in terms of 
an exponential of a quadratic form of $\bb(\z),\bc(\z)$. 
\end{abstract}
\maketitle
\bigskip

\section{ Introduction}

This paper continues our study of correlation functions
in lattice integrable models \cite{BJMST1,BJMST2,BJMST3,BJMST4,
BJMST}.
Consider the infinite XXZ spin chain  with the Hamiltonian
\begin{eqnarray}
H_{\rm XXZ}=\textstyle{\frac{1}{2}}\sum\limits_{k=-\infty}^{\infty}
\left( 
\sigma_{k}^1\sigma_{k+1}^1+
\sigma_{k}^2\sigma_{k+1}^2+
\Delta\sigma_{k}^3\sigma_{k+1}^3
\right), 
\label{eq:XXZ}
\end{eqnarray}
where $\sigma^a \, (a=1,2,3)$ are 
the Pauli matrices and
\be
\Delta=\cos\pi\nu
\en
is a real parameter. 
We use the usual notation
$$
q=e^{\pi i \nu}
\,.
$$
In our previous work \cite{BJMST}, 
we obtained an algebraic representation for
general correlation functions of the XXZ model. 
Here we generalize this result to the situation when a disorder
operator is present.
In the course we find a new interesting structure behind the model. 
We consider only the massless regime $|\Delta|<1$, $0<\nu<1$, 
since it is more important for physics because of its relation to
conformal field theory (CFT) discussed below.
Explanation about the massive regime $\Delta>1$ will be given
elsewhere.

Let us introduce 
$$S(k)=\textstyle{\frac 1 2} \sum\limits_{j=-\infty}^k\sigma ^3_j\,.$$
Denote by $|\text{vac}\rangle$ the ground state of
the Hamiltonian, and let $\al$ be a parameter. 
We consider the normalized vacuum expectation values:
\begin{align}
\frac{\langle\text{vac}|q^{2\al S(0)}
\mathcal{O}|\text{vac}\rangle}{\langle\text{vac}|q^{2\al S(0)}
|\text{vac}\rangle}\label{exp}
\end{align}
where $\mathcal{O}$ is a local operator.

Locality of $\mathcal{O}$ implies that the operator
$q^{2\al S(0)}\mathcal{O}$ stabilizes: there exist integers $k, l$ such that 
for all $j>l$ (resp. $j< k$) this operator acts on the 
$j$-th lattice site as $1$ (resp. $q^{\al\sigma ^3}$).
If $k$ (resp. $l$) is the maximal (resp. minimal) integer with this property,
$l-k+1$ will be called the length of the operator $q^{2\al S(0)}\mathcal{O}$. 
The very formulation of the problem implies that we are
interested only in local operators $\mathcal{O}$ of total spin $0$
(otherwise the correlation function vanishes). 
Nevertheless, for the sake of
convenience we introduce the spaces $\mathcal{W}
_{\al,s}$ of operators $q^{2\al S(0)}\mathcal{O}$ of
spin $s$:
$$
\[S\ ,\ q^{2\al S(0)}\mathcal{O}\]=s\ q^{2\al S(0)}\mathcal{O},
\quad S=S(\infty)\,.
$$
Also we set
$$
\mathcal{W}_{\al}=\bigoplus\limits _{s=-\infty}^{\infty}
\mathcal{W}_{\al,s}\,.
$$

The leading long distance asymptotics of the XXZ spin chain is 
described by CFT with $c=1$: 
that of free bosons $\phi,\bar{\phi}$ 
with compactification radius $\beta=\sqrt{1-\nu}$. 
{}For an extensive discussion about the XXZ model 
as an irrelevant perturbation of CFT, 
we refer the reader to \cite{luk}.  
The space $\mathcal{W}_{\al,s}$ corresponds to the space
of descendants of the operator 
$$
e^{\frac i 2\(\al (\beta ^{-1}-\beta )(\phi +\bar{\phi})+s\beta (\phi-
\bar{\phi})\)}\,.
$$
Similarly to the conformal case \cite{BLZ1,BLZ2,BLZ3}, 
introduction of the disorder parameter $\al$ 
regularizes the problem, and allows to write
much nicer formulae than in the case $\al =0$ 
\footnote{
The formulae are written initially for $|q^{\al}|<1$
and continued analytically in $\al$, but $\al =0$ is
one of singular points where l'H\^opital's rule should be applied.}.
Another similarity is that it is very
convenient to consider, as an intermediate object which
does not enter the final formulae, the following space:
$$
\mathcal{W}_
{[\al]}=\bigoplus\limits _{k=-\infty}^{\infty}\mathcal{W}_{\al+k}\ .
$$

In this paper we 
shall introduce two 
anti-commuting families of operators $\bb (\z )$ and
$\bc (\z)$ acting on $\mathcal{W}_{[\al]}$:
\begin{align}
\[\bb (\z_1),\bb (\z_2)\]_+=\[\bb (\z_1),\bc (\z_2)\]_+=\[\bc (\z_1),\bc (\z_2)\]_+=0\ .\nn
\end{align}
The operators $\bb (\z )$ and $\bc (\z )$ have the following block
structure:
$$
\bb (\z):\ \mathcal{W}_{\al+k,s}\to
\mathcal{W}_{\al+k+1,s-1}, 
\quad \bc (\z ):\ \mathcal{W}_{\al+k,s}\to 
\mathcal{W}_{\al+k-1,s+1}
\ .
$$
Hence the operator 
$\bb (\z_1)\bc (\z _2)$ acts from $\mathcal{W}_{\al,0}$ to itself.

The operators $\bb (\z )$, $\bc (\z )$ 
are formal series in $(\z -1)^{-1}$.
When applied to an operator $q^{2\al S(0)}\mathcal{O}$ of length $L$,   
the singularity is a pole of order $L$, in other words, the series terminates at
$(\z -1)^{-L}$.
The action of $\bb (\z)$, $\bc (\z)$ produces
operators of the same or smaller length. 
The coefficients of $\bb (\z)$, $\bc (\z)$ 
give rise to an action of the Grassmann algebra with $2L$
generators. In particular 
$$
\bb(\z_1)\cdots\bb(\z_{L+1})
\(q^{2\al S(0)}\mathcal{O}\)=0,
\quad 
\bc(\z_1)\cdots\bc(\z_{L+1})
\(q^{2\al S(0)}\mathcal{O}\)=0
\,.
$$

We introduce also the  linear functional
on $\text{End}\(\mathbb{C}^2\)$:
\begin{align}
\text{tr}^{\alpha}(x)=
\frac 1 {q^{\frac {\al }2}+q^{-\frac {\al }2}}
\text{tr} \(q^{-\frac 1 2 \al \sigma ^3}x\)
\label{tral}
\end{align}
with the obvious properties:
$$
\text{tr}^{\alpha}(1)=\text{tr}^{\alpha}(q^{\al \sigma ^3})=1\,.
$$
This gives rise to a
linear functional on $\mathcal{W}_{\al}$ 
$$
\mathbf{tr}^{\al}(X)=\cdots\tr _1^{\al}\ \tr _2^{\al}\ 
\tr _3^{\al}\cdots (X)\,.
$$

Our main result is:
\begin{align}
\frac{\langle\text{vac}|q^{2\al S(0)}
\mathcal{O}|\text{vac}\rangle}{\langle\text{vac}|q^{2\al S(0)}
|\text{vac}\rangle}\ =
\mathbf{tr}^{\al}\(e^{\mbox{\scriptsize\boldmath{$\Omega$}}}\(q^{2\al S(0)}\mathcal{O}\)\)\,,
\label{main}
\end{align}
where\footnote {In \cite{BJMST} the operator $\Ob$ was denoted by $\Omega ^*$.}
the operator $\Ob$ acts on $\mathcal{W}_{[\al ]}$:
\begin{align}
\Ob=-
\res _{\z_1=1}\res _{\z_2=1}
\(\ob \(\z_1/\z_2\)\bb (\z _1)\bc (\z _2)
\frac{d\z _1}{\z _1}\frac{d\z _2}{\z _2}\)\,,\nn
\end{align}
and $\ob (\z)$ is a scalar operator on each $\mathcal{W}_{\al}$,  
\begin{align}
&
\left. \ob (\z)\right|_{\mathcal{W}_{\al}}=\omega (\z ,\al)1_{\mathcal{W}_{\al }},\label{omega}
\end{align}
the scalar being
\begin{align}
&\omega (\z ,\al)\nn\\
&=\frac {4(q\z)^{\al }} {\(1+q^{\al}\)^2}
\(
\frac {q^{-\al}}{1-q^{-2}\z ^2}-\frac {q^{\al}}{1-q^2\z ^2}\)+
\int\limits _{-i\infty -0}^{i\infty -0}\z ^{u+\al}
\frac {\sin \frac {\pi} 2\(u-\nu(u+\al)\)}{\sin \frac {\pi} 2 u
\cos \frac {\pi\nu} 2\(u+\al\)}du\ \nn.
\end{align}
{}For any local operator of length $L$, 
the trace is effectively taken over $\(\mathbb{C}^2\)^{\otimes L}$.

Comments are in order about the meaning of \eqref{main}.
In \cite{JM,JM2}, 
in the setting of inhomogeneous chains,  
it was conjectured 
that the thermodynamic limit of the ground state 
averages in the finite XXZ chain 
are certain specific solutions of the reduced qKZ 
(rqKZ) equation given by multiple integrals. 
Subsequently these integral formulas were also derived 
from the viewpoint of algebraic Bethe Ansatz \cite{maillet}.   
We take these formulas as the definition of 
the left hand side of \eqref{main}.
Following our previous works \cite{BJMST2,BJMST}, 
we present here another formula for solutions of rqKZ equations. 
The right hand side of \eqref{main}
is its specialization to the homogeneous case.
We have no doubt that these two solutions coincide
\footnote{It is known to be the case in the massive regime, 
see \cite{BJMST2}. 
We also confirm the coincidence at the free fermion point, 
see section \ref{XX}.   
}. 
Since a mathematical proof is lacking at the moment,
we propose \eqref{main} as conjecture. 
The function $\omega (\z ,\al)$ and $\mathbf{tr}^{\al}$ develop singularities
at $\al = \pm 1/\nu$. 
In view of this, we presume that the formula holds true throughout the range 
$|{\rm Re}\, \al|<1/\nu$. 

It will be shown that the operators $\bb (\z)$, $\bc (\z)$ commute
with the adjoint action of the shift operator $U$ and 
of local integrals of motion $I_p$ on $\mathcal{W}_{[\al]}$. 
Since $q^{-\al S}$ commutes with $U,I_p$, 
one immediately concludes that the vacuum expectation values of 
$U\(q^{2\al S(0)}\mathcal{O}\)U^{-1}
 -q^{2\al S(0)}\mathcal{O}$ and
$\[ I_p,q^{2\al S(0)}\mathcal{O}\]$ 
given by (\ref{main}) vanish, as it should be.

In our opinion the appearance of anti-commuting operators $\bb (\z)$
and $\bc (\z)$ is quite remarkable. 
In the next section we explain how these operators are
constructed using the $q$-oscillators. 
We explain their relation to the 
Jordan-Wigner fermions in the XX case in Section \ref{XX}. 
In Appendix we briefly discuss the generalization
of our previous formulae \cite{BJMST} to the case when
the disorder operator is present. 

{}For the sake of simplicity 
we consider the homogeneous chain only. 
We give brief explanations about the inhomogeneous case when needed.
We do not give complete  proofs, but just sketch the derivation
of the main statements. 
We tried to make this paper as brief as possible,
leaving the details to a separate publication.

\section{Operators $\bb (\z )$ and $\bc (\z )$}\label{bc}

{}First we prepare our notation for the $L$-operators. 
Consider the quantum affine algebra $U_q(\slth)$. 
The universal $R$-matrix of this algebra belongs to the tensor product 
$\mathfrak{b}_+\otimes \mathfrak{b}_-$ of its two Borel subalgebras. 
By an $L$-operator we mean its image under an algebra map
$\mathfrak{b}_+\otimes \mathfrak{b}_-\to N_1\otimes N_2$, 
where $N_1,N_2$ are some algebras.  
In this paper we always take $N_2$ to be the 
algebra $M=Mat(2,\C)$ of $2\times 2$ matrices. 
As for $N_1$ we make several choices:
$U_q(\slt)$, $M$, 
the $q$-oscillator algebra 
$Osc$ (see below) or $Osc\otimes M^{\pm}$, 
where $M^{\pm}\subset M$ are the subalgebras
of upper and lower triangular matrices.
{}For economy of symbols, we use the same letter $L$ 
to designate these various $L$-operators. 
We always put indices, 
indicating to which tensor product of algebras they belong. 
We use $j,k,\cdots$ as labels for the lattice sites, 
and $a,b,\cdots$ as those for the `auxiliary' two-dimensional space. 
Accordingly we write the matrix algebra as $M_j$ or $M_a$. 
Capital letters $A,B,\cdots$ will indicate 
the $q$-oscillator algebra $Osc$.  
{}Finally, for $Osc\otimes M^{\pm}$ we use pairs of 
indices such as $\{A,a\}$.

The first case of $L$-operators is when $N_1=U_q(\mathfrak{sl}_2)$:
\begin{align}
L_j(\z)=\begin{pmatrix}\z q^{\frac{H+1}2}-\z ^{-1}q^{-\frac{H+1}2}
&(q-q^{-1})Fq^{\frac{H-1}2}\cr (q-q^{-1})q^{-\frac{H-1}2}E & \z q^{-\frac{H-1}2}-\z ^{-1}q^{\frac{H-1}2}
\end{pmatrix}_j\in U_q(\slt)\otimes
M_j.\label{Lop}
\end{align}
Here $E,F,q^{\pm H/2}$ are the standard generators of $U_q(\mathfrak{sl}_2)$. 
The suffix $j$ in the right hand side 
means that it is considered as a $2\times 2$ matrix in $M_j$. 
This is an exceptional case when we do not put any index for
the first (`auxiliary') tensor factor; we shall never use several copies 
of $U_q(\slt)$.
Mapping further $U_q(\slt)$ to $M_a$, 
we obtain the second $L$-operator 
$$
L_{a,j}(\z)\in M_a\otimes M_j\,,
$$  
which actually 
coincides with the standard $4\times 4$ $R$-matrix. 

The next case is due originally
to Bazhanov, Lukyanov and Zamolodchikov \cite{BLZ3}.
Let us consider the $q$-oscillators $a$, $a^*$ satisfying
$$ 
aa^*-q^2a^*a=1-q^2.
$$
It is convenient to introduce one more element $q^D$ such that 
\begin{align}
&q^{D}a^*=a^*
q^{D+1},\qquad q^{D}a=aq^{D-1}\,,\nn\\
&a^*a=1-q^{2D}, \quad aa^*=1-q^{2D+2}\,.\nn
\end{align}
Denote by $Osc$ the algebra generated by $a,a^*,q^{\pm D}$ 
with the above relations.
We consider the following two representations of $Osc$, 
\begin{align}
W^+=&\bigoplus\limits _{k=0}^\infty \mathbb{C}|k\rangle,
\quad a^*|k-1\rangle=|k\rangle,
\quad
D|k\rangle=k|k\rangle ,\quad a|0\rangle=0;\nn\\
W^-=&\bigoplus\limits _{k=-\infty}^{-1} \mathbb{C}|k\rangle,
\quad a|k+1\rangle=|k\rangle,
\quad
D|k\rangle=k|k\rangle ,\quad a^*|-1\rangle=0
\,.
\nn
\end{align}
In the root of unity case, if $r$ is the smallest positive 
integer such that $q^{2r}=1$, 
we consider the $r$-dimensional quotient of $W^{\pm}$ 
generated by $|0\rangle$ or $|-1\rangle$.

The $L$-operator associated with $Osc$ is given by
\begin{align}
&L^+_{A,j}(\zeta)=i\z ^{-\frac 1 2}q^{-\frac 1 4}
\begin{pmatrix}1&  -\z a_A^*\\[5pt]
 -\z  a_A &  1- \z ^2q^{2D_A+2} \end{pmatrix}_j
\begin{pmatrix}q^{D_A} &0\\[5pt] 0 & q^{-D_A}  \end{pmatrix}_j
\in 
Osc _A\otimes M_j\nn\,.
\end{align}
This $L$-operator satisfies the crossing symmetry relation:
$$
L^+_{A,j}(\z)^{-1}=\frac 1 {\zeta-\zeta ^{-1}}\overline{L}_{A,j}^+(\z)\,,
$$
where we have set
$$
\overline{L}_{A,j}^+(\z)=\sigma ^2 _j L_{A,j}^+(\z q^{-1}) ^{t _j}\sigma^2_j
\,,
$$
and $t_j$ stands for the transposition in $M_j$.
We use also another $L$-operator
$$
L^-_{A,j}(\z)=\sigma ^1_jL^+_{A,j}(\z)\sigma ^1_j\ .
$$

Consider the product $L ^+_{A,j}(\z )L_{a,j}(\z )$. 
It is well known that this product can be 
brought to a triangular form, giving rise 
in particular to Baxter's `$TQ$-equation' for transfer matrices. 
Namely, introducing
$$G ^+_{A,a}=
\begin{pmatrix} q^{-D_A} &0\\ 0& q^{D_A} \end{pmatrix}_a
\begin{pmatrix} 1& a^*_A\\ 0 &1  \end{pmatrix}_a,
\quad G^-_{A,a}=\sigma ^1_aG^+_{A,a}\sigma ^1_a\,,
$$
one easily finds that 
\begin{align}
&L^+ _{\{A,a\},j}(\z )
=\(G^+_{A,a}\)^{-1}  L^+ _{A,j}(\z)L_{a,j}(\z )G^+_{A,a}
\label{fusionright+}\\ &=
\begin{pmatrix}
(\z q-\z ^{-1}q^{-1})
L^+_{A,j}(\z q^{-1})q^{-\frac{\sigma^3_j} 2} &0\\
(q-q^{-1})L^+_{A,j}(\z q)\sigma _j^+q^{-2D_A+\frac 1 2}
& (\z -\z ^{-1})
L^+_{A,j}(\z q )q^{\frac{\sigma^3_j} 2}
\end{pmatrix}_a\in\(Osc_A\otimes M^-_a\)
\otimes M_j \,.
\nn
\end{align}

{}For the inverse matrix one has:
\begin{align}
L ^+_{\{A,a\},j}(\z )^{-1}&=\frac 1{(\z -\z ^{-1})
(\z q -\z ^{-1}q^{-1})(\z q^{-1}-\z ^{-1}q)}
\label{fusioninv}\\ &\times
\begin{pmatrix}
(\z -\z ^{-1})
q^{\frac{\sigma^3_j} 2}\ \overline{L }^+_{A,j}(\z q^{-1})&   0  \\
 -(q-q^{-1}) \sigma _j^+\ \overline{L }^+_{A,j}(\z q)q^{-2D_A+\frac 1 2}
& (\z q^{-1}-\z ^{-1}q)q^{-\frac{\sigma^3_j} 2}
\ \overline{ L }^+_{A,j}(\z   q)
\end{pmatrix}_a\,.
\nn
\end{align}
Again we shall use another $L$-operator:
$$ L ^-_{\{A,a\},j}(\z )=\sigma ^1_a\sigma ^1_jL ^+_{\{A,a\},j}(\z )
\sigma ^1_a\sigma ^1_j \in\(Osc_A\otimes M^+_a\)
\otimes M_j
\,.
$$
Some information will be needed about $R$-matrices which intertwine these 
$L$-operators.
{}First, consider the Yang-Baxter equation:
\begin{align}
&R_
{A,B}(\z_1/\z _2)L^{\pm}_{A,j}(\z _1)
L^{\pm}_{B,j}(\z _2)=L_{B,j}^{\pm}(\z _2)L^{\pm}_{A,j}(\z _1)R_{A,B}(\z_1/\z _2)
\,.
\label{YB+}
\end{align}
The $R$-matrix appearing in \eqref{YB+} is given by 
$$
R_{A,B}(\z)=P_{A,B}h(\z, u_{A,B})\zeta ^{D_A+D_B}\,,
$$
where $P_{A,B}$ is the permutation, 
$$
u_{A,B}=a_A^*q^{-2D_A} a_B,
$$
and the function $h(\z, u)$ is given by
$$
h(\z,u)=\sum\limits _{n=0}^{\infty}\ 
\(-uq^{-1}\)^n
\prod_{j=1}^n\frac{q^{j-1}\z^{-1}-q^{-j+1}\z}{q^j-q^{-j}}
\,.
$$
When $q$ is not a root of unity,   
the series for $R_{A,B}(\z)$ 
is well defined because
the action of $u_{A,B}$ on $W^{\pm}\otimes W^{\pm}$ 
is locally nilpotent.  
Otherwise we replace the right hand side 
by the sum $\sum_{n=0}^{r-1}$, 
if $r$ is the smallest positive 
integer such that $q^{2r}=1$.

Second, consider the Yang-Baxter equation for the $L$-operators
$L^+_{\{A,a\},j}$:
\begin{align}
&R ^+_{\{A,a\},\{B,b\}}(\z_1/\z _2)L^+_{\{A,a\},j}(\z _1)L^+_{\{B,b\},j}(\z _2)=
L^+_{\{B,b\},j}(\z _2)L^+_{\{A,a\},j}(\z _1)
R ^+_{\{A,a\},\{B,b\}}(\z_1/\z _2)\,.
\label{YB4}
\end{align}
The corresponding $R$-matrix has the form 
\begin{align}
&R ^+_{\{A,a\},\{B,b\}}(\zeta )=
\begin{pmatrix}
\cR_{1,1}(\z)&0 & 0 &0\\
\cR _{2,1}(\z) &\cR_{2,2}(\z)&0 & 0\\
\cR _{3,1}(\z) &0 &\cR_{3,3}(\z)&0\\
\cR _{4,1}(\z) &\cR _{4,2}(\z) &\cR_{4,3}(\z) &\cR _{4,4}(\z)
\end{pmatrix}_{a,b}
\,.
\label{R+ab}
\end{align}
The entries $\cR _{i,j}(\z )$ can be found by a direct
calculation. In this paper we shall need only two of them:
\begin{align}
&
\cR_{1,1}(\z)=q^{-D_A}R_{A,B}(\z)q^{D_B},
\quad \cR_{4,4}(\z)=-\z ^2q^{D_A}R_{A,B}(\z)q^{-D_B}\,.
\nn
\end{align}
Up to scalar coefficients depending on $\z $,  
these operators
can be guessed immediately, but the coefficient, especially
the sign, in $\cR _{4,4}(\z)$ is important for us. 
As usual we define:
$$
R ^-_{\{A,a\},\{B,b\}}(\z)
=\sigma ^1_a\sigma ^1_bR^+_{\{A,a\}\{B,b\}}(\z)
\sigma ^1_a\sigma ^1_b
\,.
$$

Now we have everything necessary for the definition of
the operators $\bb (\z )$ and $\bc (\z )$. 
{}For two integers $k\le l$ we set 
$$
M_{[k,l]}=M_{k}\otimes\cdots \otimes M_{l}\,.
$$
This is the algebra of linear operators on the `quantum space'
on the interval $[k,l]$.  
Our main object is the monodromy matrix
\begin{align}
T^{\pm}_{\{A,a\},[k,l]}(\z )=L^{\pm}_{\{A,a\},l}(\z )\cdots L^{\pm}_{\{A,a\},k}(\z )
\in Osc_A\otimes M^{\mp}_a\otimes M_{[k,l]}\,.
\label{monodromy0}
\end{align}
Define further an element
$\bT ^{\pm}_{\{A,a\},[k,l]}(\z)\in 
Osc_A\otimes M^{\mp}_a\otimes \text{End}(M_{[k,l]})$
by setting
\begin{align}
&\bT ^{\pm}_{\{A,a\},[k,l]}(\z)(X_{[k,l]})= 
T^{\pm}_{\{A,a\},[k,l]}(\z )\cdot 
(1_{A,a}\otimes X_{[k,l]})\cdot
T^{\pm}_{\{A,a\},[k,l]}(\z )^{-1}
\, ,
\label{monodromy}
\end{align}
where $1_{A,a}=1_A\otimes 1_a$ and $X_{[k,l]}\in M_{[k,l]}$. 
To illustrate the definition, we have, for 
$x_{\{A,a\}}\in Osc _A\otimes M_a^{\mp}$ and $X_{[k,l]}\in
M_{[k,l]}$, an equality in 
$Osc_A\otimes M^{\mp}_a\otimes M_{[k,l]}$
\begin{align}
&\(\bT ^{\pm}_{\{A,a\},[k,l]}(\z )\cdot x_{\{A,a\}}\otimes id\)\(X_{[k,l]}\)\nn\\
&=T^{\pm}_{\{A,a\},[k,l]}(\z )\cdot\(1_{\{A,a\}}\otimes X_{[k,l]}\)\cdot
T^{\pm}_{\{A,a\},[k,l]}(\z )^{-1}\cdot 
(x_{\{A,a\}}\otimes 1_{[k,l]})\,,\nn
\end{align}
where 
$id$ is the identity operator in $\text{End}(M_{[k,l]})$.

We define $\bT^{\pm}_{A,[k,l]}(\z)\in Osc_A\otimes \text{End}(M_{[k,l]})$ and $\bT_{a,[k,l]}(\z)\in M_a\otimes \text{End}(M_{[k,l]})$ in a similar manner.

In the following we shall use only $\bT ^{\pm}_{\{A,a\},[k,l]}(\z)^{-1}$. 
We understand certain inconvenience in using the inverse
operators, but it has for us a 
historical reason:
once we define the transfer-matrix as in \cite{JM}, the order
of multipliers is fixed everywhere.

We have the Yang-Baxter equation 
\begin{align}
&\bT ^{\pm}_{\{A,a\},[k,l]}(\z_1)^{-1}\bT ^{\pm}_{\{B,b\},[k,l]}(\z_2)^{-1}
R^{\pm}_{\{A,a\},\{B,b\}}(\z_1/\z _2)
\label{rightYB}
\\&\quad=
R^{\pm}_{\{A,a\},\{B,b\}}(\z_1/\z _2)
\bT ^{\pm}_{\{B,b\},[k,l]}(\z_2)^{-1}
\bT ^{\pm}_{\{A,a\},[k,l]}(\z_1)^{-1}\,, 
\nn
\end{align}
where the identity is in $Osc_A\otimes M^{\mp}_a\otimes
Osc_B\otimes M^{\mp}_b\otimes \text{End}(M_{[k,l]})$. 

Of particular importance are the $Q$-operators acting on local operators. 
They are defined as 
\begin{align}
& \mathbf{Q} ^+_{[k,l]}(\z, \al )=
\tr ^+_A\(q^{2\al D_A}\ \bT^{+}_{A,[k,l]}(\z)^{-1}\)(1-q^{2(\al-\bS)}), 
\label{Qop}\\
 & 
\mathbf{Q}^-_{[k,l]}(\z, \al )=\tr ^-_A\(q^{-2\al (D_A+1)}\ \bT ^{-}_{A,[k,l]}(\z)
 ^{-1}\)q^{2\bS }
 (1-q^{2(\al-\bS)})\,,
\nn
\end{align}
where $\mathbf {S} $ stands for the adjoint action of the total spin
operator 
$$ 
\mathbf {S} (X_{[k,l]})=\bigl[
S(l)-S(k-1)\ ,\ X_{[k,l]}\bigr]\, ,
\qquad
X_{[k,l]}\in M_{[k,l]}\,.
$$
The trace functionals $\tr ^{+}_A\bigl(q^{2\al D_A}Y_A\bigr)$ and 
$\tr ^{-}_A\bigl(q^{-2\al (D_A+1)}Y_A\bigr)$ for $Y_A\in Osc_A$ are defined
as analytic 
continuations with respect to $\al$ of traces over
$W^+$ and $W^-$ from the region $|q^{\al}|<1$. 
The $Q$-operators \eqref{Qop} are mutually commuting families 
of operators. They are
so normalized that $\mathbf{Q}^{\pm}_{[k,l]}(0,\al)=1$.

Regarding $\bT ^{\pm}_{\{A,a\},[k,l]}(\z)^{-1}$ 
as a matrix in $M^{\mp }_a$, let us write its entries as
\begin{align}
&\bT ^{+}_{\{A,a\},[k,l]}(\z)^{-1}=\begin{pmatrix}
\bbA ^+_{A,[k,l]}(\z)&0\\
\bbC ^+_{A,[k,l]}(\z)&\bbD ^+_{A,[k,l]}(\z)
\end{pmatrix}_a,\nn\\
&\bT ^{-}_{\{A,a\},[k,l]}(\z)^{-1}=\begin{pmatrix}
\bbA ^-_{A,[k,l]}(\z)&\bbB ^-_{A,[k,l]}(\z)\\
0&\bbD ^-_{A,[k,l]}(\z)
\end{pmatrix}_a\, ,
\nn
\end{align}
where $\bbA ^+_{A,[k,l]}(\z)$, etc., are elements of
$Osc_A\otimes \text{End}(M_{[k,l]})$.
It follows from the definition that $\bT ^{\pm\ }_{\{A,a\},[k,l]}(\z)^{-1}$ 
have poles of order $l-k+1$ at the points 
$\z^2=1, q^{\pm 2}$. 
However, looking at the formulae (\ref{fusionright+})--
(\ref{fusioninv}), one realizes that 
at the pole $\z ^2=1$ only $\bbC ^+_{A,[k,l]}(\z)$ and
$\bbB ^-_{A,[k,l]}(\z)$ are singular.  
This motivates, at least partly, the following definition:
\begin{align}
&\bc _{[k,l]}(\z,\al)=q^{\al -\bS }\(1-q^{2(\al-\bS)}\)
\text{sing}_{\,\z=1}\left[
\z ^{\al -\mathbf{S}}\tr ^+_A\(q^{2\al D_A}\ \bbC ^+_{A,[k,l]}(\z)
\)
\right],\label{defc1}\\
&\bb _{[k,l]}(\z,\al)={q^{2\bS}} \text{\,sing}_{\,\z=1}\left[
\z ^{-\al +\mathbf{S}}\tr ^-_A\(q^{-2\al(D_A+1)}\ \bbB ^-_{A,[k,l]}(\z)
\)
\right]\,.
\label{defb1}
\end{align}
Here and after, $\text{\,sing}_{\z=1}[f(\z )]$ signifies the singular part of 
$f(\z)$ at $\z =1$:
\begin{align}
\text{\,sing}_{\z=1}\[f(\z)\]=
\frac 1 {2\pi i}\int \frac { f(\xi)}  {\z-\xi}d\xi
\,,
\label{int}
\end{align}
where the integral is taken over a simple closed curve containing 
$\xi=1$ inside, while 
$\xi =\z $ and other singular points of $f(\xi)$ are outside. 
We note that 
$$
[\bS,\bc _{[k,l]}(\z,\al)]=\bc _{[k,l]}(\z,\al),
\quad 
[\bS,\bb _{[k,l]}(\z,\al)]=-\bb _{[k,l]}(\z,\al)\,.
$$

There are several important properties of operators $\bc _{[k,l]}(\z,\al)$ and
$\bb _{[k,l]}(\z,\al)$ which we formulate as Lemmas.
\begin{lem}\label{lem1}
The operators $\bc _{[k,l]}(\z,\al)$ and
$\bb _{[k,l]}(\z,\al)$ satisfy the following anti-commutation relations:
\begin{align}
&\bc _{[k,l]}(\z_1,\al-1)\bc _{[k,l]}(\z _2,\al)=-\bc _{[k,l]}(\z _2,\al-1)\bc _{[k,l]}(\z_1,\al)
\,,
\label{commcc}\\
&\bb _{[k,l]}(\z_1,\al+1)\bb _{[k,l]}(\z _2,\al)=-\bb _{[k,l]}(\z _2,\al+1)\bb _{[k,l]}(\z_1,\al)\,.
\label{commbb}
\end{align}
\end{lem}

\begin{proof}
Consider the Yang-Baxter equations (\ref{rightYB}) for $+$. Using the $R$-matrix (\ref{R+ab}) one finds:
\begin{align}
&\z_1^{-2}\bbC^+_{A,[k,l]}(\z _1)\bbC^+_{B,[k,l]}(\z _2)
\label{CC}\\
&+{\z _2} ^{-2}q^{D_A}R_{A,B}(\z_1/\z_2)q^{-D_B}\cdot
\bbC^+_{B,[k,l]}(\z _2)\bbC ^+_{A,[k,l]}(\z _1)\cdot
q^{-D_B}R_{A,B}(\z_1/\z_2)^{-1}q^{D_A}=\cdots\nn
\end{align}
where $\cdots $ stands for a sum of 
terms which contain at least one
$\bbA^+_{[k,l]}(\z _i)$ or $\bbD ^+_{[k,l]}(\z _i)$,  
and hence have vanishing singular parts at $\z _i=1$.
Multiplying (\ref{CC}) by $q^{2(\al-1)D_A+2\al D_B}$, 
taking the trace and the singular part, 
one immediately gets (\ref{commcc}).
Similarly one proves (\ref{commbb}) using (\ref{rightYB}) for $-$.
\end{proof}

\begin{lem}\label{lem2}
We have the following reduction relations:
\begin{align}
&\bc _{[k,l]}(\z ,\al)\(X_{[k,l-1]}\cdot 1_l\)=\bc _{[k,l-1]}(\z ,\al )\(X_{[k,l-1]}\)\cdot 1_l\,,
\label{redc+}\\
&\bb _{[k,l]}(\z ,\al)\(X_{[k,l-1]}\cdot 1_l\)=\bb _{[k,l-1]}(\z ,\al )\(X_{[k,l-1]}\)\cdot 1_l\,,
\label{redb+}\\
&\bc _{[k,l]}(\z ,\al)\(q^{\al\sigma ^3_k}\cdot X_{[k+1,l]}\)=
q^{(\al-1)\sigma ^3_k}\cdot
\bc _{[k+1,l]}(\z ,\al)\(X_{[k+1,l]}\)
\,,
\label{redc-}\\
&\bb _{[k,l]}(\z ,\al)\(q^{\al\sigma ^3_k}\cdot X_{[k+1,l]}\)=
q^{(\al+1)\sigma ^3_k}\cdot
\bb _{[k+1,l]}(\z ,\al)\(X_{[k+1,l]}\)\label{redb-}\,.
\end{align}
\end{lem}

\begin{proof}
The equations (\ref{redc+}), (\ref{redb+}) are
trivial consequences of the definition. 
In contrast, eqs. (\ref{redc-}), (\ref{redb-})
are far from being obvious. 

Consider the first of them. By definition we have:
\begin{align}
&\frac 1 {q^{\al-\bS }\(1-q^{2(\al-\bS)}\)}
\bc _{[k,l]}(\z ,\al)\(q^{\al\sigma ^3_k}\cdot X_{[k+1,l]}\)\nn
\\
&=
\text{\,sing}_{\z=1}\left[
\tr ^+_A\(q^{2\al D_A}\ \bbC_{A,[k,l]}^+(\z)\(q^{\al\sigma ^3_k}\cdot X_{[k+1,l]}\)
\)\z^{\al -s-1}
\right]\,,
\nn
\end{align}
where $s$ is the spin of $X_{[k+1,l]}$.

Let us simplify the trace.  
We will use the crossing symmetry
\begin{align}
&\cP ^- _{j,\bar j}
L ^+_{A,j}(\z q ^{-1})L ^+_{A, \bar j}(\z )=
(\z -\z^{-1})\cP ^- _{j,\bar j}\,,
\label{cross1}\\
&\cP ^- _{j,\bar j}
L ^+_{\{A,a\},j}(\z q ^{-1})L ^+_{\{A,a\}, \bar j}(\z )=
(\z q -\z^{-1}q^{-1})(\z -\z^{-1})(\z q^{-1}-\z^{-1}q)\cP ^- _{j,\bar j}
\,,
\label{cross2}
\end{align}
where $\cP^-_{j,\bar j}$ is the anti-symmetrizer. 
Introducing consecutively some additional two-dimensional spaces, we have 
\begin{align}
&\tr ^+_A\(q^{2\al D_A}\ \bbC_{A,[k,l]}^+(\z)\(q^{\al\sigma ^3_k}\cdot X_{[k+1,l]}\)\)
\label{trtrtr}
\\
&=
\tr _a\tr ^+_A\(\sigma _a^+L_{\{A,a\},k}^+(\z)^{-1}\cdot
\bT_{\{A,a\},[k+1,l]}^+(\z)^{-1}\(X_{[k+1,l]}\)\cdot
q^{\al\sigma ^3_k} L_{\{A,a\},k}^+(\z)q^{2\al D_A}\)\nn\\&=
\frac {1}{(\z -\z ^{-1})
(\z q -\z ^{-1}q^{-1})(\z q^{-1}-\z ^{-1}q)}\nn\\&\times
\tr _{\bar k}\tr _a\tr ^+_A\(\sigma _a^+L_{\{A,a\},\bar k}^+(\z q^{-1})
\cdot
2\cP^-_{k,\bar k}\cdot
\bT_{\{A,a\},[k+1,l]}^+(\z)^{-1}
\(X_{[k+1,l]}\)\cdot
q^{\al\sigma ^3_k} 
\cdot
L_{\{A,a\},k}^+(\z)q^{2\al D_A}\)\nn\,.
\end{align}
Now use
$$
q^{\al \sigma ^3_k}L ^+_{A,k}(\z)q^{2\al D_A}
=q^{2\al D_A}L^+_{A,k}(\z)q^{\al \sigma ^3_k}
$$ 
and the cyclicity of the trace to simplify (\ref{trtrtr}) further:
\begin{align}
&\tr ^+_A\(q^{2\al D_A}\ \bbC_{A,[k,l]}^+(\z)\(q^{\al\sigma ^3_k}
\cdot X_{[k+1,l]}\)\)
\label{exp1}
\\
&=
\frac {1}{(\z -\z ^{-1})
(\z q -\z ^{-1}q^{-1})(\z q^{-1}-\z ^{-1}q)}
\nn
\\
&\times
\tr _{\bar k}\tr _a\tr ^+_A\(
\bT_{\{A,a\},[k+1,l]}^+(\z)^{-1}\(X_{[k+1,l]}\)q^{2\al D_A}\mathcal{L}(\z )
q^{\al\sigma ^3_k} \)
\,.\nn
\end{align}
It is easy to see that
\begin{align}
\mathcal{L}(\z )&=2\cP^-_{k,\bar k}L_{\{A,a\},k}^+(\z)\sigma _a^+L_{\{A,a\},\bar k}^+(\z q^{-1})
\label{exp2}\\
&=
\begin{pmatrix}
\z q-\z^{-1}q^{-1} & 0\\
0 &(q-q^{-1})q^{-2D_A-\frac 1 2}
\end{pmatrix}_a
2\cP^-_{k,\bar k}L_{A,k}^+(\z q^{-1})L_{A,\bar k}^+(\z )
\nn\\ 
&\times
\begin{pmatrix}
q^{-\frac {\sigma ^3_k} 2} \sigma ^+_{\bar k}  &    
q^{\frac {\sigma ^3_{\bar k}-\sigma ^3_k} 2}\\[5pt]
\sigma ^+ _k \sigma ^+_{\bar k}        &      \sigma ^+ _kq^{\frac {\sigma ^3_{\bar k}} 2}
\end{pmatrix}_a
\begin{pmatrix}
(q-q^{-1})q^{-2D_A+\frac 1 2} & 0\\ 0 &\z q^{-1}-\z ^{-1}q
\end{pmatrix}_a\,.
\nn
\end{align}
where we used
$$L^+_{A,j}(\z q)\sigma _j^+q^{-2D_A+\frac 1 2}=
q^{-2D_A-\frac 1 2}L^+_{A,j}(\z q^{-1})\sigma _j^+\,.$$
In view of \eqref{cross1}, $\mathcal{L}(\z )$ is divisible by $\z-\z^{-1}$, and 
in (\ref{exp1}) we can drop the diagonal elements of 
$\bT_{\{A,a\},[k+1,l]}^+(\z)^{-1}$,  
arriving immediately at (\ref{redc-}).

The proof of (\ref{redb-}) is similar.
\end{proof}

\noindent
{\bf Remark.} 
The above construction carries over to inhomogeneous chains 
where an independent spectral parameter $\xi_j$ is attached to each site $j$. 
The operators $\bc _{[k,l]}(\z;\xi _k,\cdots ,\xi _l)$, 
$\bb _{[k,l]}(\z;\xi _k,\cdots ,\xi _l)$ 
are defined via the above construction with two modifications: 
\begin{enumerate}
\item In the definition (\ref{monodromy0}), each 
$L ^{\pm}_{\{A,a \},j}\({\z}\)$ is replaced by 
$L ^{\pm}_{\{A,a \},j}\({\z}/{\xi_j}\)$.
\item The singular part is understood as an 
integral (\ref{int}) around the points $\xi_k,\cdots,\xi_l$. 
\end{enumerate}
Lemma \ref{lem1} and Lemma \ref{lem2} remain valid. 
\hfill{\qed}
\medskip

Lemma \ref{lem2} allows us to define 
universal operators $\bb (\z,\al)$, $\bc (\z ,\al)$: 
\begin{definition}\label{defbc}
{}For any operator
$q^{2\al S(0)}\mathcal{O}\in \mathcal{W}_{\al}$, 
let 
$\(q^{2\al S(0)}\mathcal{O} \)_{[k,l]}$ be its restriction 
to the finite interval $[k,l]$ of the lattice.  
We define 
\begin{align}
&\bb (\z,\al):\ \mathcal{W}_{\al,s}\to 
\mathcal{W}_{\al+1,s-1}
\ ,\label{actb}\\
&\bc (\z ,\al ):\ \mathcal{W}_{\al,s}\to 
\mathcal{W}_{\al-1,s+1}
\label{actc}\,,
\end{align}
by setting 
\begin{align}
&\bb (\z,\al)\(q^{2\al S(0)}\mathcal{O}\)
=\lim _{k\to-\infty,l\to \infty}
\bb _{[k,l]}(\z,\al)\(\(q^{2\al S(0)}\mathcal{O}\)_{[k,l]}\)\,,
\label{defb}\\
&\bc (\z,\al)\(q^{2\al S(0)}\mathcal{O}\)
=\lim _{k\to-\infty,l\to \infty}
\bc _{[k,l]}(\z,\al)\(\(q^{2\al S(0)}\mathcal{O}\)_{[k,l]}\)\,.
\label{defc}
\end{align}
\end{definition}
It follows from Lemma \ref{lem2} that for any particular operator
$q^{2\al S(0)}\mathcal{O}$ the expressions under the limit 
in (\ref{defb}), (\ref{defc}) stabilize for sufficiently 
large interval $[k,l]$.  
Hence the limit is well-defined. 
In particular we have, for any $k$, 
\begin{align}
\bb (\z,\al)(q^{2\alpha S(k)})=0, \quad \bc (\z,\al)(q^{2\alpha S(k)})=0\,. \label{zerobc}
\end{align}

Denoting by $\bb(\z)$ and $\bc (\z)$ the operators acting on 
the direct sum $\mathcal{W}_{[\al]}$ we have 
the anti-commutativity
\begin{align}
\[\bb (\z_1),\bb (\z_2)\]_+=\[\bc (\z_1),\bc (\z_2)\]_+=0\ .\nn
\end{align}

In Appendix, we give a brief summary of 
the algebraic formula for the correlation functions in the 
presence of disorder. The result is expressed 
in terms of the operator
\begin{align}
\Ob&= -\text{res}_{\z_1=1}\text{res}_{\z_2=1}
 \(\bX(\z_1,\z _2)
 \ob (\z_2/\z _1)
 \frac {d\z _1}{\z _1} \frac {d\z _2}{\z _2}\)\,,
\label{Omega2}
\end{align}
where $\left.\bX (\z_1,\z _2)\right|_{\mathcal{W}_{\al}}=\bX (\z_1,\z _2,\al)$, the operator $\bX (\z_1,\z _2,\al)$ is  given in either of the two formulas 
(\ref{App0}), (\ref{App1}), $\ob (\z)$ is given by (\ref{omega}).  
The following result allows us to express $\Ob$ in terms of 
$\bb (\z)$, $\bc (\z )$. 
At the same time, 
the existence of two equivalent representations  
guarantees the anti-commutativity between the latter. 
\begin{lem}\label{lem3}
The operator $\bX(\z _1,\z _2)$ can be evaluated as follows:
\begin{align}
\left.\bX (\z _1,\z _2)\right|_{\mathcal{W}_{\al}}
=\bb (\z _2,\al-1)\bc (\z _1,\al)=-\bc (\z_1,\al +1)\bb (\z _2,\al)\,.
\label{X=bc=cb}
\end{align}
\end{lem}
\begin{proof}
Consider the  formula (\ref{App0}). We have:
\begin{align}
\tr _{a,b}&\(
 B^0_{b,a}(\z_2/\z_1)\bT_a(\z_1)^{-1}\bT_b(\z _2)^{-1}\)\mathbf{Q}^+(\z_1,\al+1)\mathbf{Q}^-(\z_2,\al +1)
\nn\\
&=
 \tr _{a,b}\tr _A^+\tr _B^-\(B^0_{b,a}(\z_2/\z_1)      
 \bT_a(\z_1)^{-1}\bT_b(\z _2)^{-1}
 \bT^{+}_A(\z _1)^{-1}\bT^{-}_B(\z _2)^{-1}\right. \nn
\\&\left.\times q^{2(\al+1)(D_A-D_B-1)}\)(1-q^{2(\al+1-\bS)})^2
q^{2\bS}
\,.
\nn
 \end{align}
We move $\bT_b(\z_2)^{-1}$ through $\bT^+_A(\z _1)^{-1}$ using
the Yang-Baxter equation
$$
L ^+_{A,b}\({\z _1}/{\z _2}\)\bT_b(\z_2)^{-1}\bT^{+}_A(\z _1)^{-1}
=\bT^{+}_A(\z _1)^{-1}\bT_b(\z_2)^{-1}L^+_{A,b}\({\z _1}/{\z _2}\)\,.
$$
Now $\bT_a(\z_1)^{-1} \bT^{+}_A(\z _1)^{-1}$ and 
$\bT_b(\z_2)^{-1} \bT^{-}_B(\z _2)^{-1}$  come together. 
Conjugating by $G ^+_{A,a}$, $G^-_{B,b}$,  
we can combine them into 
the monodromy matrices 
$\bT ^{+}_{\{A,a\}}(\z_1)^{-1}$, $\bT^{-}_{\{B,b\}}(\z _2)^{-1}$.
In these monodromy matrices we drop
diagonal elements 
because they have no singularities at $\z _i=1$. 
Then by a straightforward calculation we come to
\begin{align}
&\tr _{a,b}\(
 B^0_{b,a}(\z_2/\z_1)
 \bT_a(\z_1)^{-1}\bT_b(\z _2)^{-1}
\)\mathbf{Q}^+(\z_1,\al+1)\mathbf{Q}^-(\z_2,\al +1)
\label{x1}\\
&
\simeq
-
 \tr _A^+\tr _B^-\(\bbC^+_{A}(\z _1)\bbB ^-_B(\z _2)q^{2(\al+1) D_A-2\al (D_B+1)-2}\)(1-q^{2(\al-\bS+1)})^2
q^{2\bS}
\nn 
\end{align}
where $\simeq$ means that the singular parts are identical. 
Similarly we have:
\begin{align}
&\tr _{a,b}\(
 B^1_{a,b}(\z_1/\z_2)\bT_b(\z _2)^{-1}\bT_a(\z_1)^{-1}\)\mathbf{Q}^-(\z_2,\al -1)\mathbf{Q}^+(\z_1,\al-1)
\label{x2} \\
&
\simeq
- \tr _A^+\tr _B^-\(\bbB ^-_B(\z _2)\bbC^+_{A}(\z _1)
q^{2\al D_A-2(\al-1)(D_B+1)}\)
(1-q^{2(\al-\bS-1)})^2q^{2\bS}
\,.
\nn
\end{align}
Eq. (\ref{X=bc=cb}) follows from (\ref{x1}),
(\ref{x2}) and the definition of $\bb (\z ,\al)$, $\bc (\z ,\al)$.
\end{proof}
The main formula \eqref{main} follows from \eqref{Omega2},
\eqref{X=bc=cb} 
and \eqref{sol}. 

Let $U$ be the shift operator by one lattice unit, which acts 
on local operators by adjoint:
$$
U\sigma ^a_jU^{-1}=\sigma ^a_{j+1}.
$$
There is also an infinite set of local integrals of motion
which commute with $U$ and among themselves.
The last important property of $\bb (\z)$, $\bc (\z)$ 
is their invariance:
\begin{lem}\label{lem4}
The operators $\bb (\z)$, $\bc (\z)$ commute with the
adjoint action of the shift operator $U$ and of the local integrals of motion.
\end{lem}
\begin{proof}
{}For $U$ the statement of this lemma follows immediately from
the definition, essentially it is a consequence of Lemma \ref{lem2}.

The local integrals of motion are of the form
\begin{align}
I_p=\sum\limits _{j=-\infty}^{\infty} d_{j,p}\,,
\label{plocal}
\end{align}
where $d_{j,p}$ is an operator acting non-trivially on the sites 
$j,\cdots, j+p$. We shall call operators of the type (\ref{plocal}) 
$p$-local operators.

Let us write the $4\times 4$ $R$-matrix as 
$\check{R}_{j,k}(\xi)=P_{j,k}L_{j,k}(\xi)$. We set  
$$ 
U_{[k,l]}(\xi)=(q-q^{-1})^{k-l}
\check{R}_{l,l-1}(\xi)\cdots \check{R}_{k+1,k}(\xi)\,.
$$
{}Following the remark after Lemma \ref{lem2},  
consider $\bc _{[k,l]}$ with one inhomogeneity:
$$
\bc _{[k,l]}(\z;\xi, 1,\cdots, 1)\ \text{and }\ \bc _{[k,l]}(\z;1,\cdots, 1,\xi)\,.
$$
It is clear from the definition that
\begin{align}
&U_{[k,l]}(\xi)\cdot 
\bc _{[k,l]}(\z;\xi, 1,\cdots, 1)\(\(q^{2\al S(0)}\mathcal{O}\)_{[k,l]}\)\ 
\cdot U_{[k,l]}(\xi)^{-1}
\label{comm}\\
&=\bc _{[k,l]}(\z;1,\cdots, 1,\xi)
\(U_{[k,l]}(\xi)\cdot \(q^{2\al S(0)}\mathcal{O}\)_{[k,l]}\cdot 
U_{[k,l]}(\xi)^{-1}\)\,.
\nn 
\end{align}

Let $\xi=1+\epsilon$. Then 
$$
U_{[k,l]}(\xi)=\exp\(\sum\limits _{p=1}^{\infty} \epsilon^p I_{[k,l],p}\)\,.
$$
Due to the Campbell-Hausdorff formula,  
the operators $I_{[k,l],p}$ are $p$-local. 
{}For finite $k,l$ these operators do not 
commute because of some boundary terms, 
but in the limit $k\to-\infty$, 
$l\to\infty$ they coincide with the local integrals of motion 
$I_p$ which are combined into the generating function:
$$
U(\xi)=\exp\bigl(\sum\limits _{p=1}^{\infty}\epsilon ^p I_{p}\bigr)\,.
$$
 
In the right hand side of (\ref{comm}) we have the expression
$$
U_{[k,l]}(\xi)\cdot 
\(q^{2\al S(0)}\mathcal{O}\)_{[k,l]}\cdot 
U_{[k,l]}(\xi)^{-1}=\sum\limits \epsilon ^p 
\(q^{2\al S(0)}\mathcal{O}\)^{(p)}_{[k,l]}\,.
$$
Here the $p$-local operators $I_{[k,l],p}$ act by multiple adjoint.
It is clear that for every given degree $p$ we can 
find a large enough interval $[k,l]$ in order that 
$$ 
\(q^{2\al S(0)}\mathcal{O}\)^{(p)}_{[k,l]}
=\(\bigl(q^{2\al S(0)}\mathcal{O}\bigr)^{(p)}\)_{[k,l]}\,,
$$
where 
$$
U(\xi)\cdot q^{2\al S(0)}\mathcal{O}\cdot U(\xi)^{-1}=
\sum\limits \epsilon ^p \(q^{2\al S(0)}\mathcal{O}\)^{(p)}\,.
$$
Obviously
$$
\text{length}\(\bigl(q^{2\al S(0)}\mathcal{O}\bigr)^{(p)}\)\le 
\text{length}\(q^{2\al S(0)}\mathcal{O}\)+2p\,. 
$$ 
        
Now  considering (\ref{comm}) order by
order in $\epsilon$, choosing for 
every order sufficiently large interval
$[k,l]$ and using the inhomogeneous version of
Lemma \ref{lem2} and the definition of $\bc (\z)$,  
we get:
\begin{align}
&U(\xi)\cdot 
\bc (\z)\(q^{2\al S(0)}\mathcal{O}\)\cdot U(\xi)^{-1}=
\bc (\z)\(U(\xi)\cdot q^{2\al S(0)}\mathcal{O}\cdot U(\xi)^{-1}\),
\label{comm1}
\end{align}
which  is understood as an equality of power series in $\epsilon$.
\end{proof}

\section{Free fermion point}\label{XX}

Consider the point $\nu =1/2$, $q=i$. 
For this coupling constant
the Hamiltonian turns into
$$
H_{XX}=\sum\limits _{j=-\infty}^{\infty}\(\sigma _j^+\sigma _{j+1}^-+
\sigma _j^-\sigma _{j+1}^+\)\,,
$$
and can be diagonalized by the Jordan-Wigner transformation:
$$
\psi _k ^{\pm}=\sigma ^{\pm}_ke^{\mp\pi i S(k-1)
}\,.
$$
The space $\mathcal{W}_{[\al]}$ becomes a
direct sum of two components:
$$\mathcal{W}_{[\al]}=\mathcal{W}_{\al}\oplus \mathcal{W}_{\al +1}\,.$$
We set 
$$y= e^{\frac{\pi i \al} 2}\, ,$$
so that the space $\mathcal{W}_{\al}$ consists of operators of the form 
$
y^{ 2 S(0)}\mathcal{O}
$.
There are two fermion
operators acting in the space of states, so, there are 
four of them
acting on the space of operators by left and right multiplication.
It is convenient to introduce the following four operators:
\begin{align}
&\Psi _{k}^{\pm}(X)=\psi ^{\pm}_kX-(-1)^{F(X)} X\psi ^{\pm}_k,
\label{PsiPhi}\\
&\Phi_{\al,k}^{\pm}(X)=
\frac 1 {1-y^{\mp 2}}
\(\psi ^{\pm}_kX-y^{\mp 2}(-1)^{F(X)} X\psi ^{\pm}_k\)\, .
\nn
\end{align}
where $F(X)$ is the fermionic number of the operator $X$.

We have $\Phi_{\al+2,k}^{\pm}=\Phi_{\al,k}^{\pm}$. 
These operators are natural for us because $\Psi_{k}^{\pm}$
annihilate $1$ while $\Phi_{\al,k}^{\pm}$ annihilate $y^{2 S}$ 
(recall that at plus or minus infinity 
$y^{ 2 S(0)}\mathcal{O}$ stabilizes to $1$ or  $y^{2 S}$). 
The operators $\Psi_{k}^{\pm}$, $\Phi_{\al,k}^{\pm}$ satisfy the 
canonical anti-commutation relations:
\begin{align}
&[\Psi _{k}^{\epsilon},\Psi _{l}^{\epsilon '}]_+=
[\Phi_{\al,k}^{\epsilon},\Phi_{\al,l}^{\epsilon '}]_+=0,\quad
[\Psi _{k}^{\epsilon},\Phi_{\al,l}^{\epsilon '}]_+
=\delta _{\epsilon+\epsilon ',0}\delta _{k,l}\,.
\label{comferm}
\end{align}
It is clear, however, that the operators
$\bb (\z,\al)$, $\bc (\z,\al)$ cannot be constructed as linear 
combinations of $\Psi _{k}^{\pm},\ \Phi_{\al,k}^{\pm}$. 
Indeed the operators
$\bb (\z,\al)$, $\bc (\z,\al)$ are translationally invariant, in particular, they 
annihilate  $y^{2 S(k)}$ for any $k$, see (\ref{zerobc}).
Clearly this is impossible for any linear combination of 
$\Psi _{k}^{\pm},\ \Phi_{\al,k}^{\pm}$. 
Our plan in this section is as
follows. First, we find a compact expression for $\bb (\z,\al)$ and
$\bc (\z,\al)$ in terms of $\Psi _{k}^{\pm},\ \Phi_{\al,k}^{\pm}$. 
Then we show that our formula gives the same result for the correlators 
as the one 
obtained by a straightforward calculation based on normal ordering.

The calculation of $\bb (\z,\al)$, $\bc (\z ,\al)$ at the free fermon point
is summarized by
\begin{lem}\label{lem5}
At the free fermion point, the operators $\bb (\z,\al)$ 
and $\bc(\z,\al)$ are given by 
\begin{align}
&\bb (\z ,\al)=\frac{2i^{-\bS}}{1-(-1)^{\bS}y^{2}}\ 
{\rm sing}_{\z=1}
\[\z ^{-\al+\bS}\Psi^-(\z )E^{-}(\z ,\al-\bS)
\frac{\z}{1+\z^2}
\]\,,
\label{bcff}\\
&\bc (\z ,\al)=2y\ {\rm sing}_{\z=1}
\[\z ^{\al-\bS}\Psi ^+(\z )
E^{+}(\z ,\al-\bS)
\frac{\z}{1+\z^2}
\]\,,
\nn
\end{align}
where
\begin{align}
\Psi^{\pm}(\z )=\sum\limits _ {j=-\infty}^{\infty}
\Psi^{\pm}_j\(\frac {1+\z^2}
{1-\z ^2}\)^{j}\label{fermion}
\end{align}
and
\begin{align}
E^{\pm}(\z,\al)=\exp\(\mathcal{N}\[\Phi_{\al}^{\pm} 
\log \(I-\z ^2M\)
\Psi ^{\mp}-
\Phi_{\al}^{\mp} 
\log \(I+\z ^2M\)
\Psi ^{\pm}\]\)\,.\label{E}
\end{align}
In the last formula we consider $\Phi_{\al,j}^{\pm}$ 
(resp. $\Psi^{\pm}_j$)
as components of a row (resp. column) vector, 
$$ 
M=(1+u)(1-u)^{-1},\qquad 
\(u\Psi ^{\pm}\)_j=\Psi ^{\pm}_{j+1}\,,
$$
and 
$\log \(1\pm\z ^2M\)$ are understood as Taylor series in $u$.
$\mathcal{N}[\cdot]$ stands for the normal ordering 
which applies only to operators acting at the same site. For them we set
\begin{align}
\mathcal{N}[\Phi_{\al,j}^{\epsilon}\Psi_j^{\epsilon'}]
=
\begin{cases}
\Phi_{\al,j}^{\epsilon}\Psi ^{\epsilon'}_j&\quad (j>0)\,,\\
-\Psi ^{\epsilon'}_j\Phi_{\al,j}^{\epsilon}&\quad (j\le0 )\,.\\
\end{cases}
\label{norm}
\end{align}
\end{lem} 
Since the $q$-oscillators become fermions at $q=i$,    
Lemma can be shown by 
manipulations 
with exponentials of quadratic forms in fermions. 
Details will be given in another publication.

We remark that the exponent of (\ref{E}) is well defined 
as an operator on  $\mathcal{W}_{\al}$. Indeed by definition
it consists of $\mathcal{N}\(\Phi_{\al,k}^{\pm}\Psi ^{\mp}_l\)$ 
with $l\ge k$. 
On a particular operator in $\mathcal{W}_{\al}$ 
only a finite number of these operators do not vanish.
\medskip

It has been said that, unlike $\bb (\z)$, $\bc (\z)$,  
formulae containing fermions necessarily
break the translational invariance.  
We choose the point $k=1$ as the origin 
and consider only operators of the form 
\begin{align}
y^{2 S(0)}\mathcal{O}_>
\label{O+}
\end{align}
where $\mathcal{O}_> $ acts only on the interval $[1,\infty)$. 
Any operator in $\mathcal{W}_{\al}$ can be brought to the form (\ref{O+}) by a shift, 
so we do not really lose generality.
In the sequel we need the operators on a half line:
$$
\bb _>(\z,\al)=\bb_{[1,\infty)}(\z,\al),\qquad
\bc _>(\z,\al)=\bc_{[1,\infty)}(\z,\al)\,. 
$$
They are defined as in (\ref{bcff}), replacing $E^{\pm}(\z,\al)$, $\Psi ^{\pm}(\z )$ and  
$\Phi_{\al}^{\pm}(\z )$ by $E^{\pm}_>(\z,\al)$, 
$\Psi^{\pm}_>(\z )$ and  
$\Phi_{\al,>}^{\pm}(\z )$, respectively. 
The latter are  given by the same formulae (\ref{fermion}), (\ref{E}) 
with  non-positive components of fermions removed. 

In the free fermion case the function $\omega (\z ,\al)$ can be
calculated explicitly.  
Putting it together with \eqref{bcff},  
we rewrite our main formula in the free fermion case as follows. 
\begin{align}
&\frac{\langle\text{vac}|\ y^{2 S(0)}\mathcal{O}_>
\ |\text{vac}\rangle}{\langle\text{vac}|\ y^{2 S(0)}
\ |\text{vac}\rangle}
 =\mathbf{tr}_>^{\al}\(e^{\Obm_>}\(\mathcal{O}_>\)\),\label{mainferm}\\
&\Ob _>=
\frac {i} {\sin\frac {\pi  \al} 2}\res _{\z _1=1}
\res _{\z _2=1}\(
\frac {\z _1^{\al}\z_2^{-\al}-1}{\z _1^2+\z _2^2}
E^{-}_>(\z _2,\al)E^{+}_>(\z _1,\al)
\Psi_>^-(\z _2)\Psi_>^+(\z _1)
\frac{d\z_1^2}{1+\z_1^2}
\frac{d\z_2^2}{1+\z_2^2}
\),\nn
\end{align}
where $\mathbf{tr}_>^{\al}$ means that the trace is calculated
over the positive half of the chain only.

Now
notice that 
\begin{align}
&\Psi ^{\pm}_>(\z)(I)=0,
\quad \mathbf{tr}_>^{2(\al+1)}
\(\Phi_{\al,>}^{\pm}(\z)\(\mathcal{O}_>\)\)=0,\label{cran}\\
&\psi ^{\pm}_j\mathcal{O}_>
=\(\Phi_{\al,j}^{\pm}-\frac {y^{\mp 2}}{1-y^{\mp 2}}
\Psi ^{\pm}_j\)(\mathcal{O}_>)\,.
\nn
\end{align}
So, by changing $ \mathbf{tr}_>^{\al}$ to $ \mathbf{tr}_>^{2(\al+1)}$, 
the operators $\Phi_{\al,>}^{\mp}$ and $\Psi ^{\pm}_>$ can be considered
as creation-annihilation operators in the space of operators.
For efficient application of them we need 
the following:
\begin{lem}\label{lem6}
The following identity holds:
\begin{align}
\mathbf{tr}_>^{\al}\(e^{\Obm _>}\(\mathcal{O}_>\)\)=
\mathbf{tr}_>^{2(\al+1)}\(e^{\widetilde{\Obm} _>}\(\mathcal{O}_>\)\)
\label{newtrace}
\end{align}
where 
\begin{align}
\widetilde{\Ob} _>= 
\frac {i} {\sin\frac {\pi  \al} 2}\res _{\z _1=1}
\res _{\z _2=1}\(
\frac {\z _1^{\al}\z_2^{-\al}}{\z _1^2+\z _2^2}
\Psi_>^-(\z _2)\Psi_>^+(\z _1)
\frac{d\z_1^2}{1+\z_1^2}\frac{d\z_2^2}{1+\z_2^2}
\)\,.\nn
\end{align}
\end{lem}

The formulae (\ref{cran}) and (\ref{newtrace})
allow an explicit calculation of correlators. One easily obtains:
\begin{align}
\frac{\langle\text{vac}|\ y^{2 S(0)}\psi ^+_{k_1}\cdots \psi ^+_{k_p}
\psi ^-_{l_p}\cdots \psi ^-_{l_1}\ |\text{vac}\rangle }
{\langle\text{vac}|y^{2 S(0)}\ |\text{vac}\rangle }
=\text{det}\(\langle \psi ^+_{k_i} 
\psi ^-_{l_j}  \rangle\)_{i,j=1,\cdots ,p}\label{fermcorr}
\end{align}
where 
\begin{align}
\langle \psi ^+_{k} 
&\psi ^-_{l}  \rangle =\label{2point}\\
&\frac {i}{\sin \frac {\pi\al} 2}
\(-\frac {y} 2\delta _{k,l}
+\,\res _{\z_1=1}\res _{\z _2=1}
\frac {\z _1^{\al}\z _2^{-\al}}{\z _1^2+\z _2 ^2}
\(\frac {1+\z _1^2} {1-\z _1^2} \)^k\(\frac {1+\z _2^2} {1-\z _2^2} \)^l
\frac {d\z_1^2}{1+\z _1 ^2}\frac {d\z_2^2}{1+\z _2 ^2}\)\,.
\nn
\end{align}

On the other hand, 
one can calculate the correlators (\ref{fermcorr}) 
directly by normal ordering $y^{2 S(0)}$. 
The result is the same: 
(\ref{2point}) is the two-point function while (\ref{fermcorr}) is
obtained by the Wick theorem.

This calculation is unsatisfactory because
we had to pass through the fermions $\Psi^{\pm}_k$, 
$\Phi_{\al,k}^{\pm}$.
It would be much better to find a basis in the space of local
operators, on which the original
operators $\bb (\z ,\al)$, $\bc (\z ,\al)$ act nicely. 
Such a construction would have a chance to generalize to an arbitrary coupling constant.
{}For the moment we cannot do that. 

\section{Conclusion}\label{sec:conclusion}

The main result of this paper can be formulated as follows.
We consider the space $\mathcal{W}_{[\al]}$ 
of local operators in the presence of a disorder field.
We have shown that the vacuum expectation values
of operators in $\mathcal{W}_{[\al]}$ can be expressed 
in terms of two anti-commutative families of operators 
$\bb(\z)$ and $\bc(\z )$ acting on $\mathcal{W}_{[\al]}$ .
At present, we do not know 
how to organize the space $\mathcal{W}_{[\al]}$ 
in order to describe efficiently the action of $\bb(\z)$ and $\bc(\z )$.
The operators $\bb(\z)$ and $\bc(\z )$ should be considered as 
annihilation operators, as both of them kill the `vacua', i.e., operators
$q^{2\al S(k)}$, for all $k$.  
What is missing is a construction of creation operators.
Even in the free fermion case, we were able rather to
make a detour than to actually solve the problem. 

In fact, the problem of constructing creation operators
cannot be solved literally, because 
$\bb(\z )$ and $\bc(\z )$ have a large common kernel. 
Consider the restricted operators
$\bb_{[k,l]}(\z ,\al)$ and $\bc_{[k,l]}(\z ,\al )$ acting 
on the space of dimension $4^{l-k+1}$.  
In the free fermion case, it can be shown that 
the dimension of the kernel is $2^{l-k+1}$. 
Numerical experiments indicate that the dimension stays the same generically. 

Because of this kernel, we cannot expect 
operators satisfying the canonical anti-commutation relations 
with $\bb(\z )$ and $\bc(\z )$. 
So the first problem is to understand the meaning of the kernel. 
Obviously, the 
difference of any two operators in the kernel has 
vanishing expectation value. 
The origin of these operators with zero vacuum expectation 
values is a mystery to us. 
The only operators for which this property 
can be easily explained are the descendants generated
by adjoint action of local integrals of motion, 
but for them the vacuum expectation values vanish for a different 
reason: $\bb(\z)$ and $\bc(\z)$ commute with the adjoint action
of local integrals of motion as is explained by Lemma \ref{lem4}.

Understanding the origin of the kernel of $\bb(\z )$ and $\bc(\z )$,  
and the construction of creation operators, 
are the problem which we wish to solve.

\appendix
\section{}

In this appendix, we sketch how the results of \cite{BJMST}
can be modified to the situation when disorder is present. 

Consider an operator $q^{2\al S(0)}\mathcal{O}$ which is stable
on $(-\infty, k-1]$ and $[l+1,\infty)$. Consider also an 
inhomogeneous chain, 
where spectral parameters $\xi _k,\cdots ,\xi _l$ 
are attached to all the sites of the lattice where 
$q^{2\al S(0)}\mathcal{O}$ acts non-trivially. 
Then the ground state becomes dependent on $\xi _k,\cdots ,\xi _l$, and we have 
\begin{align}
&
\frac{\langle \text{vac }|\ q^{2\al S(0)}\mathcal{O}\ |\text{vac}\rangle}
{\langle \text{vac }|\ q^{2\al S(0)}|\text{vac}\rangle}
=
\mathbf{h}_{[k,l]}(\xi _k,\cdots ,\xi _l,\al)\(\(q^{2\al S(0)}\mathcal{O}\)_{[k,l]}\)\,.
\nn
\end{align}
Here $\mathbf{h}_{[k,l]}(\xi _k,\cdots ,\xi _l,\al)$ is a linear functional on
$\(\mathbb{C}^2\)^{\otimes (l-k+1)}$ 
subject to several requirements \cite{JM}:
\begin{align}
&\mathbf{h}_{[k,l]}(\xi _k,\cdots,\xi _{j+1},\xi _j ,\cdots ,\xi _l,\al)(X_{[k,l]})
\nn\\&\qquad=
\mathbf{h}_{[k,l]}(\xi _k,\cdots,\xi _j ,\xi _{j+1},\cdots ,\xi _l,\al)\(
\check{\mathbb{R}}_{j,j+1}(\xi_j/\xi _{j+1})^{-1}(X_{[k,l]})\),\nn\\
&\mathbf{h}_{[k,l]}(\xi _kq^{-1},\cdots ,\xi _l,\al)(X_{[k,l]})=\mathbf{h}_{[k,l]}(\xi _k,\cdots ,\xi _l,\al)
\(\mathbf{A }_{[k,l]}(\xi _k,\cdots ,\xi _l,\al)(X_{[k,l]})\),\nn\\
&\mathbf{h}_{[k,l]}(\xi _k,\cdots ,\xi _l,\al)\(X_{[k,l-1]}\cdot 1_l\)=
\mathbf{h}_{[k,l-1]}(\xi _k,\cdots ,\xi _{l-1},\al)\(X_{[k,l-1]}\),\nn\\
&\mathbf{h}_{[k,l]}(\xi _k,\cdots ,\xi _l,\al)\(q^{\al \sigma ^3_k}\cdot X_{[k+1,l]}\)=
\mathbf{h}_{[k+1,l]}(\xi _k,\cdots ,\xi _l,\al)\( X_{[k+1,l]}\)
\,,
\nn
\end{align}
where
\begin{align}
&\check{\mathbb{R}}_{j,j+1}(\xi_j/\xi_{j+1})(X)=
\check{R}_{j,j+1}(\xi_j/\xi_{j+1})
X
\bigl(\check{R}_{j,j+1}(\xi_j/\xi_{j+1})\bigr)^{-1},
\nn
\\
&
\mathbf{A }_{[k,l]}(\xi _k,\cdots ,\xi _l,\al)(X)
=
\left((T^{-1})^{t_k}\cdot
\sigma^2_k\cdot X q^{-\al \sigma^3_k}\cdot \sigma^2_k\right)^{t_k}
\cdot T \,,
\nn\\
&T=R_{k,l}(\xi_k/\xi_l)\cdots R_{k,k+1}(\xi_k/\xi_{k+1})\,.
\nn
\end{align}
The way for solving 
these equations is absolutely parallel to the one 
described in \cite{BJMST}. 
The answer takes the form 
\begin{align}
\mathbf{h}_{[k,l]}(\xi _k,\cdots ,\xi _l,\al)\(X_{[k,l]}\)=\mathbf{tr}^{\al}
\(e^{\Obm _{[k,l]}(\xi _k,\cdots ,\xi _l,\al)}\(X_{[k,l]}\)\)
\,.
\label{sol}
\end{align}
In order to describe $\Ob _{[k,l]}$, 
let $\bT_{[k,l]}(\z)\in U_q(\slt)\otimes \text{End}(M_{[k,l]})$ denote  the monodromy matrix 
(\ref{monodromy}) constructed via the $L$-operator (\ref{Lop}).
(Normally we do not write the dependence on $\xi _k,\cdots ,\xi _l$
explicitly.)  

Introduce operators 
$
\bX^i_{[k,l]}(\z _1,\z _2,\al)$ ($i=0,1$) 
depending rationally on $\z_1,\z_2$ 
by setting 
\begin{align}
&\tr _{a,b}\(B^0_{a,b}(\z _1/\z _2)\bT_{b,[k,l]}(\z _2)^{-1}\bT_{a,[k,l]}(\z _1)^{-1}\)
\Tr _{d(\z _1/\z_2)}\(\bT_{[k,l]}\(\sqrt{\z _1\z _2}\)^{-1}q^{-(\al+1)H}\)
\nn\\
&=\(\frac {\z_1}{\z _2}\)^{\al}\bX^0_{[k,l]}(\z _1,\z _2,\al)+
\(\frac {\z_2}{\z _1}\)^{\al}\bX^0_{[k,l]}(\z _2,\z _1,\al),\nn\\
&\tr _{a,b}\(B^1_{a,b}(\z _1/\z _2)\bT_{b,[k,l]}(\z _2)^{-1}\bT_{a,[k,l]}(\z _1)^{-1}\)
\Tr _{d(\z _1/\z_2)}\(\bT_{[k,l]}\(\sqrt{\z _1\z _2}\)^{-1}q^{-(\al-1)H}\)
\nn\\
&=\(\frac {\z_1}{\z _2}\)^{\al}\bX^1_{[k,l]}(\z _1,\z _2,\al)+
\(\frac {\z_2}{\z _1}\)^{\al}\bX^1_{[k,l]}(\z _2,\z _1,\al)\,.\nn
\end{align}
Here $\Tr_{d}$ stands for the trace functional 
on $U_q(\mathfrak{sl}_2)$ 
(analytic continuation of the trace with respect 
to dimension $d$) used in \cite{BJMST2}, 
$$
d(\z)=\frac{\log \z}{\log q}\,,
$$
and $B^0(\z)$, $B^1 (\z ) $ are $4\times 4$ matrices given by 
$$B^0(\z)=\frac {(\z-\z^{-1})}{(\z q-\z ^{-1}q^{-1})(\z q^{-1}-\z q)}
 \begin{pmatrix}
 0 &0 &0 &0\\
 0 &q & -\z ^{-1} &0\\
 0  & -\z  &  q^{-1} &0\\
 0 &0 &0 &0
\end{pmatrix}$$
and 
$$
B^1(\z)=\(\sigma^1\otimes\sigma^1\)\cdot
B^0(\z)\cdot\(\sigma^1\otimes\sigma^1\)\,.
$$  

The notation being as above, the formula for $\Ob_{[k,l]}$ is 
given as follows. 
\begin{align}
\Ob_{[k,l]}(\xi _k,\cdots ,\xi _l,\al)&= 
-\frac 1 {2\pi i}\int\!\!\int
\mathbf{X}_{[k,l]}(\z_1,\z _2,\al)
\omega (\z_2/\z _1,\al)
\frac {d\z _1}{\z _1} \frac {d\z _2}{\z _2}
\label{Omega}
\end{align}
where integrals are taken around $\xi _k,\cdots ,\xi _l$.  
The operator $\bX _{[k,l]}(\z _1,\z _2,\al )$ is 
presented in either of the following two equivalent forms:
\begin{eqnarray*}
\bX_{[k,l]}(\z_1,\z _2,\al)
&=&
\text{\,sing}_{\z _1,\z_2=\xi_k,\cdots,\xi_l}
\[
\bX_{[k,l]}^0(\z _1,\z _2,\al)
\(\frac{\z_1}{\z_2}\)^{\al-\bS}
 \]
\\
&=&
\text{\,sing}_{\z _1,\z_2=\xi_k,\cdots,\xi_l}
\[
\bX_{[k,l]}^1(\z _1,\z _2,\al)  
\(\frac{\z_1}{\z_2}\)^{\al-\bS}
\]\,.
\nn
\end{eqnarray*}
In the formulas \eqref{sol}, \eqref{Omega}
only operators of spin $0$ are considered. Here we have introduced
$\bS$ for later convenience. 
 
Now we give the only part of the construction which has no analogues
in \cite{BJMST}. 
Let us consider the homogeneous case.
In particular, we replace $\frac 1 {(2\pi i)^2}\int\int$ 
by $\text{res}_{\z_1=1}\text{res}_{\z_2=1}$ 
and drop the index $[k,l]$.
{}From \cite{BLZ3} we learn:
\begin{align}
&\Tr _{d(\z _1/\z_2)}\(\bT\(\sqrt{\z _1\z _2}\)^{-1}q^{-\al H}\)
=
\nn\\ &=
\(\(\frac {\z_1}{\z _2}\)^{\al}\bQ^+(\z_1,\al)\bQ^-(\z_2,\al )-
\(\frac {\z_2}{\z _1}\)^{\al}\bQ^-(\z_1,\al)\bQ^+(\z_2,\al )\)
\frac {1} {q^{\al -\bS}-q^{-\al +\bS}}\,.
\nn
\end{align}
This implies 
\begin{align}
\bX(\z _1,\z _2,\al)&\nn\\
&=
\text{\,sing} _{\z _1=1}\text{\,sing} _{\z _2=1}\[
\(\frac {\z _1}{\z _2}\)^{\al+1-\bS}
\tr _{a,b}\(
B^0_{b,a}(\z_2/\z_1)\bT_a(\z_1)^{-1}\bT_b(\z _2)^{-1}\)\right.\label{App0}\\&\times\left.
 \bQ^+(\z_1,\al+1)\bQ^-(\z_2,\al +1)\hskip -.5cm{\ _{\ }\atop \ _{\ }}^{\ }\right]
\frac{-1}{q^{\al+1-\bS}-q^{-\al-1+\bS}}
\nn\\
&=
\text{\,sing} _{\z _1=1}\text{\,sing} _{\z _2=1}
\[
\(\frac {\z_1}{\z_2}\)^{\al-1-\bS}
\tr _{a,b}\(
 B^1_{a,b}(\z_1/\z_2) \bT_b(\z _2)^{-1}\bT_a(\z_1)^{-1}\)\right.\label{App1}\\&\times\left.\bQ^-(\z_2,\al -1) \bQ^+(\z_1,\al -1)\hskip -.5cm{\ _{\ }\atop \ _{\ }}^{\ }\right]
\frac {1}{q^{\al-1-\bS}-q^{-\al+1+\bS}}\,.
\nn
\end{align}
These two formulae will be used in Section 2 to derive a new 
expression for $\Ob$ (see Lemma \ref{lem3}).
\vskip .5cm

\noindent
{\it Acknowledgements.}\quad
Research of HB is supported 
by the RFFI grant \#04-01-00352.
Research of MJ is 
supported by 
the Grant-in-Aid for Scientific Research B--18340035.
Research of TM is
supported by 
the Grant-in-Aid for Scientific Research B--17340038.
Research of 
FS is supported by INTAS grant \#03-51-3350, by EC networks  "EUCLID",
contract number HPRN-CT-2002-00325 and "ENIGMA",
contract number MRTN-CT-2004-5652
and GIMP program (ANR), contract number
ANR-05-BLAN-0029-01. 
Research of YT is supported by the Grant-in-Aid for Young Scientists B--17740089. 
This work was also supported by the grant of 21st Century 
COE Program at RIMS, Kyoto University. 
The authors would like to thank 
Boris Feigin and Leon Takhtajan for interest and 
discussions. HB is also grateful to Rodney Baxter,
Frank G{\"o}hmann, Andreas Kl{\"u}mper,
Ingo Peschel and Junji Suzuki for discussions.  
HB would like to thank Department of Mathematics, Graduate
School of Science, Kyoto University for warm hospitality.
MJ is grateful to Christian Korff and the 
staff members of the City University of London 
for kind invitation and hospitality, where a 
part of this work has been carried out. 
FS is grateful to RIMS, Kyoto University for
always warm and nostalgic atmosphere.

\end{document}